\documentclass[twocolumn,english,amssymb,aps,prb]{revtex4}
\usepackage[T1]{fontenc}
\usepackage[latin9]{inputenc}
\usepackage{units}
\usepackage{amstext}
\usepackage{graphicx}
\usepackage{amssymb}

\makeatletter

\providecommand*{\perispomeni}{\char126}
\AtBeginDocument{\DeclareRobustCommand{\greektext}{%
  \fontencoding{LGR}\selectfont\def\encodingdefault{LGR}%
  \renewcommand{\~}{\perispomeni}%
}}
\DeclareRobustCommand{\textgreek}[1]{\leavevmode{\greektext #1}}
\DeclareFontEncoding{LGR}{}{}

\newcommand{\lyxmathsym}[1]{\ifmmode\begingroup\def\b@ld{bold}
  \text{\ifx\math@version\b@ld\bfseries\fi#1}\endgroup\else#1\fi}

\@ifundefined{textcolor}{}
{%
 \definecolor{BLACK}{gray}{0}
 \definecolor{WHITE}{gray}{1}
 \definecolor{RED}{rgb}{1,0,0}
 \definecolor{GREEN}{rgb}{0,1,0}
 \definecolor{BLUE}{rgb}{0,0,1}
 \definecolor{CYAN}{cmyk}{1,0,0,0}
 \definecolor{MAGENTA}{cmyk}{0,1,0,0}
 \definecolor{YELLOW}{cmyk}{0,0,1,0}
 }

\makeatother

\usepackage{babel}

\begin{document}

\title{Finite Conductivity in Mesoscopic Hall Bars of Inverted InAs/GaSb
Quantum Wells }

\author{Ivan Knez}

\email{ik5@rice.edu}

\affiliation{Department of Physics and Astronomy, Rice University, Houston, TX
77251-1892}

\author{R.R. Du}

\affiliation{Department of Physics and Astronomy, Rice University, Houston, TX
77251-1892}

\author{Gerard Sullivan}

\affiliation{Teledyne Scientific and Imaging, Thousand Oaks, CA}
\begin{abstract}
We have studied experimentally the low temperature conductivity of
mesoscopic size InAs/GaSb quantum well Hall bar devices in the inverted
regime. Using a pair of electrostatic gates we were able to move the
Fermi level into the electron-hole hybridization state, and observe
a mini gap. Temperature dependence of the conductivity in the gap
shows residual conductivity, which can be consistently explained by
the contributions from the free as well as the hybridized carriers
in the presence of impurity scattering, as proposed by Naveh and Laikhtman
{[}Euro. Phys. Lett., 55, 545-551 (2001){]}. Experimental implications
for the stability of proposed helical edge states will be discussed. 
\end{abstract}
\maketitle


It was recently proposed by Kane and Mele\cite{1}, and independently,
by Bernevig and Zhang\cite{2} that a novel transport phenomenon,
termed the quantum spin Hall effect (QSHE) should arise in a class
of two-dimensional topological insulators (TI). The QSHE phase is
characterized by an odd number of Kramer pair states at the edges,
and an insulating gap in the bulk. Edge states are robust against
disorder due to the topology of the bands and result in helical edge
transport with counter-propagating spin-up and spin-down channels.
Four-probe transport measurements on such structures at zero magnetic
field give a near quantized conductance of $2e^{2}/h$ for mesoscopic
samples. Just as theoretically predicted by Bernevig et al\cite{3},
the QSHE was soon observed in HgTe/CdTe quantum wells\cite{4}. More
recently, Liu et al\cite{5} proposed to observe QSHE in an inverted
InAs/GaSb composite quantum well (CQW), in which the band structure
can be tuned with electrical fields\cite{6}, giving rise to a phase
transition from a normal insulator to TI via a continuously varying
parameter. 

In CQW, carriers (electrons and holes) are separated in different
wells and the bulk energy gap arises due to the hybridization of the
e- and the h- bands at a finite $k$-vector in momentum space. This
finite $k$ character is in contrast to that in the HgTe QWs, where
the carriers are in the same well and the energy gap opens in the
zone center. It is theoretically realized that disorder (which always
exists in real materials) could play different roles in the two cases\cite{7}
for the experimental observation of QSHE. In this paper we present
an experimental study of low-temperature conductivity in mesoscopic
size CQW Hall-bar devices. Analyses of our results indicate intricate
contributions from the free as well as the hybridized carriers to
the bulk CQW conductivity in the presence of a finite amount of disorder.
Interestingly, such contributions appear to be independent of the
amount of scattering but vary strongly with CQW band parameters\cite{8}. 

Transport studies of CQW were reported early by Yang et al\cite{9}
and by Cooper et al\cite{10}. In {[}9{]} a mini-gap originating from
hybridization was observed in capacitance measurements and the density
of states (DOS) in the gap regime was analyzed. In {[}10{]}, the longitudinal
resistance $R_{xx}$ of a larger size CQW Hall bar was measured, observing
distinct maxima in $R_{xx}$ as a function of gate voltage, as the
Fermi energy moves through an energy gap at the anti-crossing points
of the e- and h- dispersions. Far-infrared measurements\cite{11,12}
corroborated the existence of a tunneling-induced mini-gap. These
results convincingly show that a mini-gap exists due to e-h hybridization
in the inverted regime of InAs/GaSb CQWs. In contrast, a true bulk
insulator, which shows temperature activated conductivity has yet
to be experimentally confirmed. Moreover, in light of the current
interest in observing the QSHE, transport in the mesoscopic-size devices
should be measured\cite{4}. Work reported here presents a first experimental
study in this direction. 

InAs/GaSb CQW was grown by molecular beam epitaxy on silicon-doped
$\mathrm{N}^{+}$(100) GaAs substrate. The structure consists of a
standard buffer consisting of AlSb and $\mathrm{Al_{08}Ga_{02}Sb}$
layers which accommodates for about 7\% lattice mismatch between GaAs
and AlSb\cite{13}. On top of this a 500Å AlSb lower barrier was grown,
followed by 150Å InAs and 80Å GaSb quantum wells with a 500Å AlSb
top barrier and 30Å GaSb cap layer to prevent from oxidation of the
AlSb barrier. The sample in which our main results were measured is
a Hall bar with width and length of 0.7\textgreek{m}m$\times$1.5\textgreek{m}m
(see inset of FIG. 1c), processed using standard photo- and e-beam
lithography techniques with wet etching. The top gate was fabricated
by depositing 2500Å Si$_{3}$N$_{4}$ using plasma enhanced chemical
vapor deposition system, and evaporating 1000Å Al or Ti/Au metal gate.
$\mathrm{N}^{+}$GaAs substrate serves as a back gate for our devices
and was contacted using silver resin. Ohmic contacts to the electron-hole
layers were made with indium and without annealing. Low temperature
magnetotransport measurements were carried out in a $^{3}$He refrigerator
(300 mK) combined with a 12T superconducting magnet, or in a $^{3}$He/$^{4}$He
dilution refrigerator (20 mK) with a 18T magnet (National High Magnetic
Field Laboratory). Standard lock-in technique with an excitation current
of 100nA at 23Hz was employed. 

\begin{figure}
\includegraphics{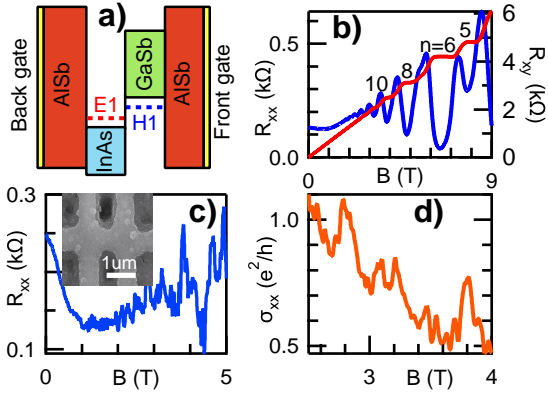}
\caption{\label{FIG 1} {(Color online) a) Shows structure and energy spectrum
of inverted CQW with E1<H1. Separation of the bands, $E_{go}$, as
well as Fermi energy $E_{F}$ can be tuned with front and back gates.
When $E_{F}$ >H1, only electrons are present in the well and representative
magneto-transport data at $T=\unit[0.3]{K}$ is shown in b) for 10\textgreek{m}m$\times$20\textgreek{m}m,
and in c) for 0.7\textgreek{m}m$\times$1.5\textgreek{m}m Hall bar
(SEM image in the inset) where $R_{xx}$ exhibits strong fluctuations.
d) Fluctuations in conductivity are on the order of $e^{2}/h$, indicating
mesoscopic regime.} }

\end{figure}

The energy spectrum of CQW is schematically shown in FIG. 1a. Due
to the broken gap alignment of InAs and GaSb, conduction and valence
states are confined in InAs and GaSb layers, respectively\cite{14}.
For wider wells such as ours, the structure is inverted (FIG. 1a),
with the ground conduction subband (E1) lower than the ground heavy-hole
subband (H1), resulting in anti-crossing of the bands at a finite
momentum value $k_{cross}$(FIG. 2a). Assuming that band anisotropy
is small, anti-crossing occurs when carrier densities in two wells
are equal, $n=p=k_{cross}^{2}/2\pi$, corresponding to the resonant
condition of equal particle energy and in-plane momentum in two wells.
Due to the tunneling induced coupling, electron and hole states are
mixed and a mini-gap $\lyxmathsym{\textgreek{D}}$ opens in otherwise
semi-metallic band dispersion (FIG. 2a)\cite{6}. 

\begin{figure}
\includegraphics{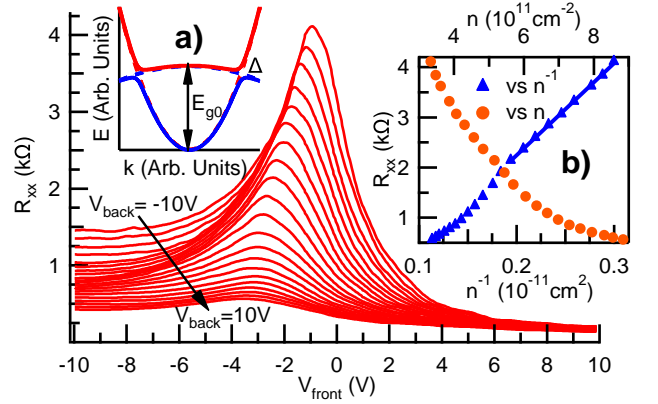}
\caption{\label{FIG 2} {(Color online)Shows $R_{xx}$ vs. $V_{front}$ for
$V_{back}$ from 10 to -10 V, in 1V steps, $B=\unit[0]{T}$, $T=\unit[0.3]{K}$.
a) At the anti-crossing point, where $n=p=E_{g0}\frac{m^{*}}{\pi\hbar^{2}}$,
a hybridization gap $\lyxmathsym{\textgreek{D}}$ opens. For $E_{F}$
in the gap, $R_{xx}$ exhibits resonance peaks, which decrease for
increasing $V_{back}$. b) Resonance peaks vary lineary with $n^{-1}$for
$n\lesssim5\cdot10^{11}\mathrm{cm^{-2}}$as proposed in {[}8{]}.} }

\end{figure}

In dual gate geometry (FIG 1a), both the relative separation between
the subbands, $E_{g0}$, and the Fermi energy $E_{F}$ can be tuned.
When $E_{F}$ is between H1 and E1, electrons and holes coexist in
their respective layers, resulting in two-carrier transport with characteristic
non-linear dependence of Hall resistance on magnetic field\cite{9}.
Single carrier transport occurs for $E_{F}$ above H1 or below E1,
and is electron- or hole-like, respectively. In our CQW, $E_{F}$
is pinned by the surface states in the cap layer\cite{15} and under
zero applied bias only electrons are present in the well, with a typical
low temperature density of $7\cdot10^{11}\mathrm{cm^{-2}}$ and mobility
of $\unit[90,000]{cm^{2}/Vs}.$ Shubnikov de Haas (SdH) oscillations
can be observed starting at 1.8T with no evidence of parallel conduction.
Hall resistance varies linearly with magnetic field until the appearance
of the integer quantum Hall plateaus. A representative trace for electron-like
transport in a larger size Hall bar (10\textgreek{m}m$\times$20\textgreek{m}m)
is shown in FIG. 1b. In contrast, micron-size devices show strong
fluctuations in $R_{xx}$ (FIG. 1c). Fluctuations are reproducible
in magnetic field, and conductivity varies on the order of $e^{2}/h$
(FIG. 1d), indicating mesoscopic regime. 

In the single carrier regime, electron density changes linearly with
front and back bias as approximately $1.5\cdot10^{11}\mathrm{cm^{-2}/V}$
and $0.4\cdot10^{11}\mathrm{cm^{-2}/V}$, respectively. Electron densities
are extracted by sweeping gate biases at fixed magnetic field or through
analysis of SdH oscillations. Values are consistent and agree well
with the parallel capacitor model. When holes are induced in the GaSb
well, the effect of front gate on the electron density in the InAs
layer is near-perfectly screened. When induced, holes will coexist
with electrons in the device bias range and hole density can be extracted
by fitting $R_{xy}$ with two-carrier transport expression. Obtained
values are consistent with those for unscreened electrons and in agreement
with the parallel capacitor model. 

In FIG. 2 we sweep the front bias from 10 to -10V with the back bias
fixed, changing the carriers from solely electrons to predominantly
holes, as evidenced from the change of sign in Hall resistance at
high magnetic fields. When the carrier densities are matched (i.e.,
$n\sim p$), clear peaks in $R_{xx}$ appear, indicating the existence
of a mini-gap. Resonance peaks, $R_{xx}(max)$, increase with decreasing
back bias, and hence vary inversely with resonance electron-hole density,
$n=p$ and corresponding $k_{cross}$(FIG. 2). In particular, for
$n\lesssim\unit[5\cdot10^{11}]{\mathrm{cm^{-2}}}$, resonance peaks
vary linearly with $n^{-1}$ (FIG. 2b). This inverse relationship,
which we subsequently discuss, cannot be explained with an increasing
mini-gap, for coupling between wells varies proportionally with $k$\cite{16}.

\begin{figure}
\includegraphics{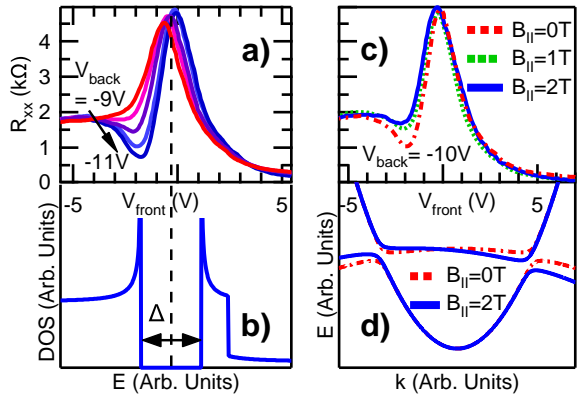}
\caption{\label{FIG 3} {(Color online) a) Shows $R_{xx}$ vs. $V_{front}$
for $V_{back}$ from -9 to -11V, in $0\unit[.5]{V}$ steps, $B=\unit[0]{T}$,
$T=\unit[20]{mK}$. Resistance dips occur at singularities in DOS
near gap edges shown in b). From relative position of dips and peaks
in $V_{front}$ we determine $\lyxmathsym{\textgreek{D}}\approx\unit[3.6]{meV}$.
Dips weaken with in-plane magnetic field in c), due to induced anisotropy
in the dispersion in d).} }

\end{figure}

Our central finding is the existence of finite conductivity in InAs/GaAs
in the mini-gap regime. Even at the lowest temperature $T=\unit[20]{mK}$
(FIG. 3a), largest observed resonance peaks $R_{xx}(max)$ remain
$\leq\unit[5]{k\text{\textgreek{W}}}$ for mesoscopic samples, which
is at least 2-3 times smaller than $h/2e^{2}$. This is in contrast
to the case of HgTe QWs where quantized value is approached from larger
resistance values\cite{4}. Hence, observed resonance peaks can be
understood as a bulk effect, with a residual conductivity on the order
of $10e^{2}/h$ per square, which is a few times larger than the predicted
contribution from the edge. For the remainder of this paper we concentrate
on the origin of residual gap conductivity and discuss its implications
for the proposed edge modes. We first note that the anisotropy of
the heavy hole band may play a role in the gap anisotropy at the Fermi
energy, which could lead to residual conductivity. Anisotropy is more
apparent for larger $k_{cross}$\cite{16} and may explain the decrease
of $R_{xx}(max)$ with larger $n$ (FIG. 2). To what extent such anisotropy
would affect the gap value is a subject for numerical calculations
with realistic materials parameters\cite{17}.

Here we show experimentally that band anisotropy plays a minor role
in our transport regime. Measurements at $T=\unit[20]{mK}$ reveal
clear dips in $R_{xx}$ in the vicinity of resonance peaks, for $V_{back}<\unit[\mathrm{-9}]{V}$
(FIG. 3a). Such regime corresponds to small on-resonance carrier densities,
and observed dips can be explained by van der Hoove singularities
in DOS (FIG. 3b) at gap edges\cite{16}. In fact, this is the first
time that such singularities have been observed in transport, attesting
to the high quality of our samples. Resistance dips occur only for
smaller $k$ due to otherwise increasing anisotropy of the valence
band, smearing out DOS singularities. Applying in-plane magnetic field
(FIG. 3d) induces anisotropy in the direction of the field\cite{9}
and dips weaken (FIG. 3c). Thus, at least in the small carrier density
regime band anisotropy cannot be responsible for the observed residual
conductivity.

\begin{figure}
\includegraphics{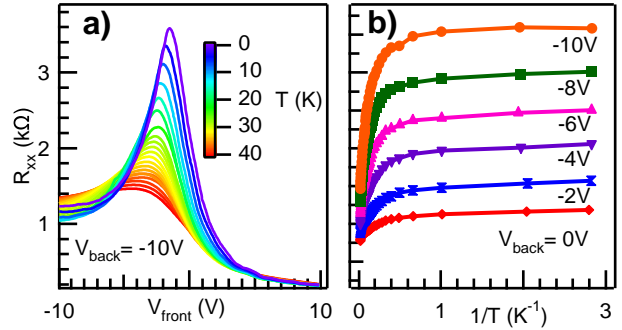}
\caption{\label{FIG 4} {(Color online) a) Shows temperature dependence of
resonance peaks for $V_{back}$=$\unit[-10]{V}$ for $T$ from $\unit[0.3]{K}$
to $\unit[40]{K}$ in roughly $\unit[2]{K}$ steps. b) Resonance peaks
shown vs $T^{-1}$ saturate for $T<\unit[2]{K}$, for six different
$k_{cross}$.}}

\end{figure}

We estimate the gap value from the relative position of resistance
dips and peaks (FIG. 3a, 3b) in front gate bias as: $\triangle=2\left(V_{peak}-V_{dip}\right)\frac{\Delta p}{\Delta V}\frac{1}{DOS},$
where $\frac{\Delta p}{\Delta V}$ is the rate of carrier density
change with front bias, and $DOS=(m_{e}+m_{h})/\pi\hbar^{2},$ with
carrier masses $m_{e}=0.03$ and $m_{h}=0.37$ (in units of free electron
mass)\cite{9}. We obtain $\triangle=\unit[3.6]{meV}$, in agreement
with previous studies\cite{9,10,12}. Similar value can be deduced
from temperature dependence of resonance peaks (FIG. 4a), which shows
insulating character, i.e. $R_{xx}(max)$ increases in lowering $T$.
On the other hand, we do not observe temperature-activated resistance,
as one would expect for a true insulator. Resonance peaks increase
only by a factor of 2-3 over three orders of magnitude change in temperature,
for six anti-crossing points measured, and saturate for $T<\unit[2]{K}$
(Fig. 4b). Nevertheless, peaks persist up to 40 K for all six cases,
consistent with a gap value of 3-4 meV.

The lack of gap activation and large residual conductivity suggest
that disorder may be important in this system. Hence, we analyze carrier
scattering times characterizing the transport in a zero magnetic field.
At zero bias, electron scattering time is $\tau_{r}=\unit[1.5]{ps}$
and the associated level broadening is $\Gamma_{e}=\hbar/2\tau_{r}=\unit[0.2]{meV}$.
Electron mobility shows linear dependence on electron density and
drops to approximately $\unit[40,000]{cm^{2}/Vs}$ for the case of
smallest electron density of $3.3\cdot10^{11}\mathrm{cm^{-2}},$ giving
$\Gamma_{e}=\unit[0.5]{meV}$. Linear dependence of mobility on electron
density is indicative of short range scattering, presumably from dislocations
at interfaces, which is confirmed by extracting the quantum time $\tau_{q}$
from SdH oscillations\cite{18}, giving $\tau_{r}/\tau_{q}\text{\ensuremath{\approx}}7$
at zero bias and validating predominance of large angle scattering.
Estimated $\tau_{q}$ has been corrected for density inhomogeneity\cite{19}
with Gaussian width of $\lyxmathsym{\textgreek{d}}n\approx0.35\cdot10^{11}\mathrm{cm^{-2}}$.
On a side note, this suggests that the random potential fluctuations
are on the order of $\text{\textgreek{d}}E=\text{\textgreek{d}}n/DOS\sim\unit[0.25]{meV}$,
and thus $\text{\textgreek{d}}E\ll\Delta$. For holes, $\Gamma_{h}=b\frac{m_{e}}{m_{h}}\Gamma_{e}$,
where $b$ is the ratio of electron to hole mobilities. From a fit
to $R_{xy}$ in two-carrier transport regime, $b\approx6$, giving
$\Gamma_{h}=\unit[0.3]{meV}$. Thus, total level broadening is less
than $\Gamma=\Gamma_{e}+\Gamma_{h}\approx\unit[0.8]{meV}$, which
is a few times smaller than the size of the gap, indicating that disorder
may play non-trivial role. 

Laikhthman and Naveh\cite{8} have theoretically studied the transport
in the inverted regime of InAs/GaAs system, concluding that even negligible
but finite level broadening due to the carrier scattering will result
in finite on-resonance conductivity at $T=\unit[0]{K}$. Specifically,
residual conductivity will go as $\sigma_{on}\left(T=0\right)\sim\frac{e^{2}}{h}\frac{E_{g0}}{\Delta}$,
when $\Gamma\ll\Delta\ll E_{g0}$, thus independent of scattering
parameters. Since $E_{g0}=n\frac{\pi\hbar^{2}}{m^{*}}$, where $m*=\frac{m_{e}m_{h}}{m_{e}+m_{h}}$,
it follows that $R_{xx}(max)\sim n^{-1}$ as observed in FIG. 2b for
$n\lesssim5\cdot10^{11}\mathrm{cm^{-2}}$. 

The physics of the unusual result in {[}8{]} can be understood by
re-examining the original premise for carrier hybridization, which
is the non-locality of electron states in growth direction of the
wells\cite{20}. If this non-locality is suppressed, then hybridization
and associated transport properties will be suppressed as well. A
natural way for this to happen is through localization of electrons
and holes in their respective wells by scattering. Contribution to
residual conductivity from such scattered carriers can be quantitatively
described within the relaxation time model. The relevant time-scale
is carrier tunneling time, $\tau_{t}=\hbar/2\Delta$, and total number
of scattered carriers within $\tau_{t}$ will go as $n\cdot(1-e^{-\frac{\tau_{t}}{\tau_{r}}})$,
or equivalently as $n\cdot(1-e^{-\frac{\Gamma}{\Delta}})$. Scattered
carriers do not hybridize and contribute to residual conductivity
as $\sigma_{on}\approx\frac{e^{2}}{h}\frac{E_{g0}}{\Gamma}\left(1-e^{-\frac{\Gamma}{\Delta}}\right)$,
using that electron mobility is $\mu_{e}=\frac{e}{m}\frac{\hbar}{2\Gamma}$
and ignoring smaller contribution from scattered holes. In the limit
$\Gamma\ll\Delta,$ we recover $\sigma_{on}\approx\frac{e^{2}}{h}\frac{E_{g0}}{\Delta}$
. Thus, because the number of non-hybridized carriers goes as $\lyxmathsym{\textgreek{G}}$
and their mobility goes as $1/\lyxmathsym{\textgreek{G}}$, it follows
that the bulk conductivity, which is their product, will have constant
value independent of $\lyxmathsym{\textgreek{G}}$. 

Besides the contribution to gap conductivity from scattered electrons
and holes, Ref. {[}8{]} also suggests an \textquotedblleft{}unusual\textquotedblright{}
contribution from hybridized electron-holes, which can be understood
as the consequence of non-zero charge of hybridized states due to
level broadening. In a two band calculation using independent e- and
h- states as a basis\cite{21}, expectation value of this charge can
be shown to go as $\Gamma/\Delta$ for $\Gamma\ll\Delta$. Since the
mobility varies as $1/\lyxmathsym{\textgreek{G}}$ and the density
of hybridized e-h changes slightly, we obtain $\sigma_{on}\approx\frac{2e^{2}}{h}\frac{E_{g0}}{\Delta}$
, where the factor of 2 comes from two types of hybridized states.
Total conductivity is then $\sigma_{on}\approx\frac{3e^{2}}{h}\frac{E_{g0}}{\Delta}$
and from the slope of the linear fit in FIG. 2b we extract the value
of the hybridization gap, obtaining $\lyxmathsym{\textgreek{D}}\approx\mathrm{\unit[4]{meV}}$
\textendash{} surprisingly close to previous estimates considering
the simplicity of the model. 

In conclusion, our experiment confirmed the existence of a hybridization-induced
energy gap in inverted InAs/GaSb. However, we observed finite bulk
conductivity in the gap regime, which can be reasonably explained
by the model proposed in {[}8{]}. Since the edge modes are only protected
by an insulating gap, presence of states inside the bulk gap will
allow for scattering of edge states between the opposite sides of
the sample, destroying helical transport properties. Nevertheless,
an interesting regime for QSHE could exist at the critical point,
where the band structure changes from inverted to normal. In contrast
to the inverted case, in the normal regime, a real insulating gap
exists, and in the vicinity of the critical point, carrier masses
can still have opposite signs\cite{6} to those in vacuum due to band
repulsion, giving rise to helical edge states. Our work in this parameter
regime is in progress. 

We thank S. -C. Zhang and D.C. Tsui for bringing our attention to
QSHE in InAs/GaAs system, H. Kroemer for invaluable advices on materials,
J. Kono, X. -L. Qi, and Kai Chang for very helpful discussions. The
work at Rice was supported by Welch Foundation Grant Number C-1682,
and a Rice Faculty Initiative Fund. Travel to NHMFL was supported
by NSF Grant No. DMR-0706634. A portion of this work was performed
at the National High Magnetic Field Laboratory, which is supported
by NSF Cooperative Agreement No. DMR-0084173, by the State of Florida,
and by the DOE. We thank T. P. Murphy and J. H. Park for expert technical
assistance. 

\end{document}